\documentclass[preprint,aps,a4paper,showpacs,amsfonts,amsmath,amssymb]{revtex4}
\usepackage{makeidx}
\usepackage{graphicx}
\usepackage{epsfig}

\begin{document}

\title{The escape problem under stochastic volatility: the Heston model} 

\author{Jaume Masoliver}
\email{jaume.masoliver@ub.edu}
\affiliation{Departament de F\'{\i}sica Fonamental, Universitat de Barcelona,\\
Diagonal, 647, E-08028 Barcelona, Spain}
\author{Josep Perell\'o}
\email{josep.perello@ub.edu}
\affiliation{Departament de F\'{\i}sica Fonamental, Universitat de Barcelona,\\
Diagonal, 647, E-08028 Barcelona, Spain}

\date{\today}

\begin{abstract}
We solve the escape problem for the Heston random diffusion model. We obtain exact expressions for the survival probability (which ammounts to solving the complete escape problem) as well as for the mean exit time. We also average the volatility in order to work out the problem for the return alone regardless volatility. We look over these results in terms of the dimensionless normal level of volatility --a ratio of the three parameters that appear in the Heston model-- and analyze their form in several assymptotic limits. Thus, for instance, we show that the mean exit time grows quadratically with large spans while for small spans the growth is systematically slower depending on the value of the  normal level. We compare our results with those of the Wiener process and show that the assumption of stochastic volatility, in an apparent paradoxical way, increases survival and prolongs the escape time.

\end{abstract}
\maketitle

\section{Introduction}
\label{sec1}

Models of financial dynamics based on two-dimensional diffusion processes, known as stochastic volatility (SV) models \cite{ronnie_book}, are being widely accepted as a reasonable explanation for many empirical observations collected under the name of ``stylized facts'' \cite{cont}. In such models the volatility, that is, the standard deviation of returns, originally thought to be a constant, is a random process coupled with the return so that they both form a two-dimensional diffusion process governed by a pair of Langevin equations \cite{ronnie_book}.

Volatility is nowadays a key magnitude in any financial setting. It is the backbone of many financial products that are designed to cover investors' risk. Extreme values associated with volatility have thus a special meaning, as they do in physics and natural sciences where escape problems in noisy environments such as Kramers problem are of the utmost importance \cite{risken,redner}. 

In a recent paper we have addressed a partial aspect of the problem: that of extreme times for the volatility regardless the value of the return \cite{mp}. Now we want to address the overall escape problem associated with both return and volatility. This is certainly a more difficult task  because the return strongly depends on volatility while, in the standard approach to SV models, the latter is supposed to be independent of the former. 

We are thus left with a two-dimensional escape problem which is always quite involved. The situation is similar to that of the unbounded Brownian particle where the extreme-value problem for the velocity of the particle is relatively easy to handle while that of its position is much more intricate \cite{risken,masoliver,mpo_prl,mpo_pre}. 

The extreme-time problem of the return has been addressed, to our knowledge, in only few works. We refer the reader to our recent work on the subject \cite{mmp,montero_lillo}, although it is based on the continuous-time random walk technique which is an entirely different frame, yet with a different scope, than that of SV models. Within the setting of the latter we are only aware of the recent work of Bonano et al \cite{spagnolo1,spagnolo2} in which an approach to the escape-time problem is addressed through simulations of the Heston model. 

In this paper we study the complete extreme value problem of one particular SV model: the Heston model \cite{heston}. Different SV models basically differ in the way the volatility depends on the underlying noise governing its dynamics. The Heston model has the benefit, over other SV models, of allowing exact analytical developments. This is the case of its unrestricted ({\it i.e.,} barrier-free) probability density function which was obtained by Yakovenko and Dragulescu few years ago \cite{yakov}. Herein we will obtain not only the exact expression of the the mean escape time (MET) but the exact survival probability as well. Being the knowledge of the latter equivalent to solving the entire escape problem. 

The paper is organized as follows. In Sec. \ref{sec2} we present the Heston model and obtain the complete solution to the escape problem. In Sec. \ref{sec3} we evaluate the mean escape time and analyze its behavior for high and low volatility. In Sec. \ref{sec4} we average out the volatility assuming it has reached the stationary state. This allows us to get exact expressions for the survival probability and the mean escape time of the return alone. Conclusions are drawn in Sec. \ref{sec5} and some more technical details are in appendices.

\section{The Heston model and the survival probability}
\label{sec2}

Let $P(t)$ be a speculative price or the value of a financial index. We define the zero-mean return $X(t)$ through the stochastic differential:
\begin{equation}
dX(t)=\frac{dP(t)}{P(t)}-\left\langle\frac{dP(t)}{P(t)}\right\rangle,
\label{zero-mean}
\end{equation}
where $\langle\cdot\rangle$ denotes the average. In terms of $X(t)$ the Heston model \cite{heston} is a two-dimensional diffusion process $(X(t),Y(t))$ described by the following pair of stochastic differential equations 
\begin{equation}
dX(t)=\sqrt{Y(t)}dW_1(t),
\label{dx}
\end{equation}
\begin{equation}
dY(t)=-\alpha\left[Y(t)-m^2\right]dt+k\sqrt{Y(t)}dW_2(t),
\label{dy}
\end{equation}
where $W_i(t)$ are Wiener processes, i.e. $dW_i(t)=\xi_i(t)dt$ $(i=1,2)$, where $\xi_i(t)$ are zero-mean Gaussian white noises with $\langle\xi_i(t)\xi_i(t')\rangle=\delta_{ij}\delta(t-t')$ \cite{footnote1}. Note that in this particular  model the volatility is 
\begin{equation}
\sigma(t)=\sqrt{Y(t)},
\label{volatility}
\end{equation}
i.e., $Y(t)$ is the variance of return although, as long as no confusion arises, we will use the term ``volatility variable'' or just ``volatility'' for the random process $Y(t)$. In Eq. (\ref{dy}) the parameter $m$ is the so-called normal level of volatility,  $\alpha>0$ is related to the ``reverting force'' toward the normal level (see below) and $k$, sometimes referred to as the ``vol-of-vol'', measures the fluctuations of the volatility. 

In the context of biological diffusion problems the process $Y(t)$ described by Eq.~(\ref{dy}) was proposed many years ago by Feller~\cite{feller} who, among other properties, proved that $Y(t)$ is always positive so that the volatility,  Eq. (\ref{volatility}), is real, positive and well defined. This feature along with a non-negligible (and exponential) autocorrelation with characteristic time $1/\alpha$ makes the process very appealing from the  perspective of mathematical finance. 

In 1985 Cox, Ingersoll and Ross~\cite{cox} introduced the same dynamics in connection with interest rates of bonds. Almost a decade later and aiming to provide a more realistic price for options, Heston \cite{heston} undertook the same dynamics but for the diffusion coefficient of financial price fluctuations as is precisely shown in Eqs. (\ref{dx})-(\ref{dy}). 

The resulting process has become quite popular among financial practitioners who want to include the effect of volatility changes in option pricing. Part of this success is due to the easy interpretation of the parameters. As mentioned, $1/\alpha$ provides the typical time that the volatility needs to reach the stationary state (the stationary density is the Gamma distribution, see Sec. \ref{sec4}). For this reason, $\alpha$ can also be interpreted as the strength of the reverting force that ties the process $Y(t)$ to its normal level $m^2$, the latter being the mean value of $Y(t)$ in the stationary state. The magnitude of the volatility fluctuations is provided by $k$ which like $\alpha$ and $m^2$ have all units of $1/({\rm time})$.

Our main interest is the escape problem associated with the Heston model. To this end, let us denote by $S(x,y,t)$ the probability that the zero-mean return $X(t)$, starting at $X(0)=x$ with volatility $Y(0)=y$, is at time $t$ inside the interval $(-L/2,L/2)$ without having ever left it during previous times. In other words, $S(x,y,t)$ is the survival probability (SP) for the joint process $(X(t), Y(t))$ to be at time $t$ inside the strip
$$
-L/2\leq X(t)\leq L/2,  \qquad 0<Y(t)<\infty,
$$
with $X(0)=x$ and $Y(0)=y$. 

The SP obeys the following backward Fokker-Planck equation \cite{gardiner}
\begin{equation}
\frac{\partial S}{\partial t}=-\alpha(y-m^2)\frac{\partial S}{\partial y}+
\frac{1}{2}k^2y\frac{\partial^2 S}{\partial y^2}+\frac{1}{2}y\frac{\partial^2 S}{\partial x^2},
\label{fpe}
\end{equation}
with initial and boundary conditions respectively given by
\begin{equation}
S(x,y,0)=1,\qquad S(\pm L/2,y,t)=0.
\label{ibc}
\end{equation}

This problem can be solved by means of Fourier series. Indeed the boundary conditions, $S(\pm L/2,y,t)=0$, lead us to look for a solution of the form
\begin{equation}
S(x,y,t)=\sum_{n=0}^\infty S_n(y,t)\cos\left[\left(2n+1\right)\pi x/L\right],
\label{fss}
\end{equation}
where the Fourier coefficients $S_n(y,t)$ are
\begin{equation}
S_n(y,t)=\frac{2}{L}\int_{-L/2}^{L/2} S(x,y,t)\cos\left[\left(2n+1\right)\pi x/L\right]dx.
\label{fcs}
\end{equation}
From Eqs. (\ref{fpe}) and (\ref{fcs}) we see that these coefficients are the solution to the initial-value problem
\begin{equation}
\frac{\partial S_n}{\partial t}=-\alpha(y-m^2)\frac{\partial S_n}{\partial y}+
\frac{1}{2}k^2y\frac{\partial^2 S_n}{\partial y^2}-
\frac{1}{2}\left[(2n+1)\pi/L\right]^2 y S_n,
\label{fpe1}
\end{equation}
with initial condition
\begin{equation}
S_n(y,0)=\gamma_n,
\label{ic}
\end{equation}
where
\begin{equation}
\gamma_n=\frac{4(-1)^n}{\pi(2n+1)}.
\label{gamma}
\end{equation}

Defining a new time scale $\tau$ and a new volatility variable $v$ by the change of scale
\begin{equation}
\tau=\alpha t, \qquad v=(2\alpha/k^2)y, 
\label{tau-v}
\end{equation}
the problem above reads
\begin{equation}
\frac{\partial S_n}{\partial\tau}=-(v-\theta)\frac{\partial S_n}{\partial v}+
v\frac{\partial^2 S_n}{\partial v^2}-(\beta_n/2L)^2 v S_n,
\label{fpe-new}
\end{equation}
and
\begin{equation}
S_n(v,0)=\gamma_n,
\label{ic-new}
\end{equation}
where
\begin{equation}
\beta_n=(k/\alpha)(2n+1)\pi,
\label{beta}
\end{equation}
and
\begin{equation}
\theta=(2\alpha/k^2)m^2.
\label{theta}
\end{equation}

Before proceeding further let us remark that the parameter $\theta$, which turns out to be crucial for the escape problem at hand, can be regarded as the ``dimensionless normal level'' of volatility. It represents a balance between the tendency toward the normal level measured by $\alpha m^2$ and the volatility fluctuations quantified by $k^2$ (see discussion in Sec. \ref{subsec4b} regarding the cases $\theta<1$ and $\theta>1$).

The problem posed by Eqs. (\ref{fpe-new})-(\ref{ic-new}) is solved by the function
\begin{equation}
S_n(v,\tau)=\gamma_n\exp\left\{-A_n(\tau)-B_n(\tau)v\right\},
\label{solution_1}
\end{equation}
where $A_n(\tau)$ and $B_n(\tau)$ are functions of time to be determined. In effect, plugging it into Eq. (\ref{fpe-new}) we see that Eq. (\ref{solution_1}) is the solution to the problem provided that 
\begin{equation}
A_n(\tau)=\theta\int_0^\tau B_n(s) ds,
\label{A_def}
\end{equation}
and $B_n(\tau)$ obeys the Riccatti equation
\begin{equation}
\dot{B_n}=-B_n-B_n^2+(\beta_n/2L)^2,
\label{riccatti}
\end{equation}
with initial condition $B_n(0)=0$. 

In the Appendix \ref{appA} we show that 
\begin{equation}
A_n(\tau)=\theta\left[\mu_-\tau+\ln\left(\frac{\mu_++\mu_-e^{-\Delta_n\tau}}{\Delta_n}\right)\right],
\label{A}
\end{equation}
and
\begin{equation}
B_n(\tau)=\mu_-\frac{1-e^{-\Delta_n\tau}}{1+(\mu_-/\mu_+)e^{-\Delta_n\tau}},
\label{B}
\end{equation}
where
\begin{equation}
\Delta_n=\sqrt{1+(\beta_n/L)^2},\qquad\qquad \mu_{\pm}=(\Delta_n\pm 1)/2.
\label{delta}
\end{equation}

\begin{figure}
\epsfig{file=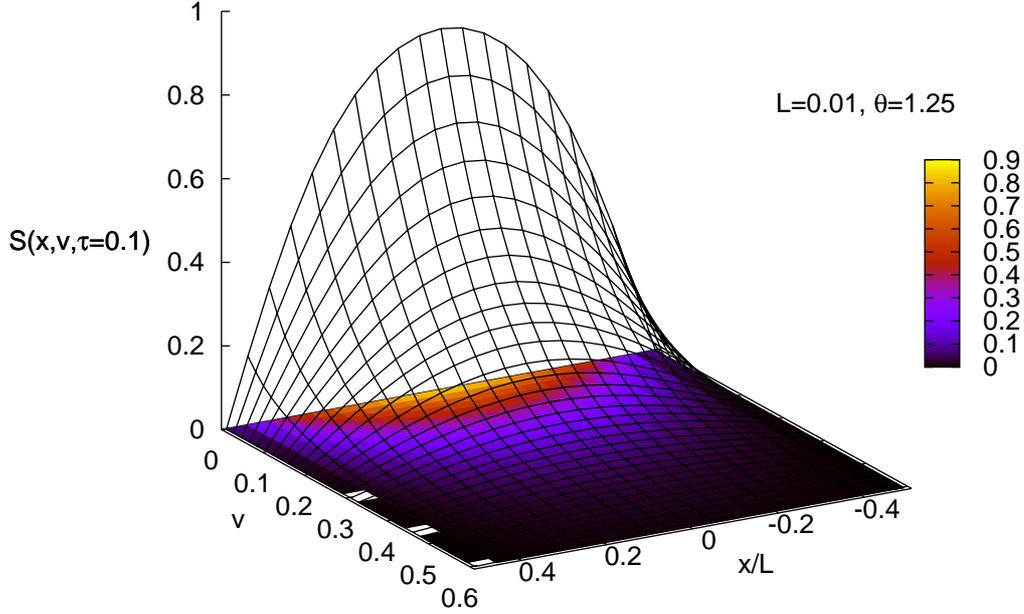}
\caption{(Color online) The survival probability $S(x,v,\tau)$ given by Eq.~(\ref{complete_solution}) with $\tau=0.1$ 
($t=2.22 \mbox{ days}$) and $L=0.01$ in terms of return $x$ and volatility $v$. Parameters of the model: $\theta=1.25$, $\alpha=0.045 \mbox{ day}^{-1}$, $m=0.093 \mbox{ day}^{-1/2}$ and $k=0.0014 \mbox{ day}^{-1}$. Recall Eq. (\ref{theta}) and note that there only exist three independent parameters.}
\label{sp}
\end{figure}

Therefore the solution to the escape problem for the two-dimensional Heston SV model is
\begin{equation}
S(x,v,\tau)=\sum_{n=0}^\infty \gamma_n\exp\left\{-A_n(\tau)-B_n(\tau)v\right\}
\cos\left[\left(2n+1\right)\pi x/L\right].
\label{complete_solution}
\end{equation}
Figure~\ref{sp} shows in a three-dimensional plot this SP as a function of the return $x$ and volatility variable $v$ for $\tau=0.1$ \cite{footnote2}.

In the asymptotic regime, either for long or short times, the SP is somewhat simpler. Thus when $\tau\gg 1$ (i.e., $t\gg\alpha^{-1}$) we have $A_n(\tau)\sim\theta\mu_-\tau$ and $B_n(\tau)\sim\mu_-$. Hence
\begin{equation}
S(x,v,\tau)\simeq \sum_{n=0}^\infty e^{-\mu_-(\theta\tau+v)}
\cos\left[\left(2n+1\right)\pi x/L\right],\qquad (\tau\gg 1).
\label{fsg_long}
\end{equation}

On the other hand for short times $\tau\ll 1$ (i.e., $t\ll\alpha^{-1}$) we write 
$e^{-\Delta_nt}=1-\Delta_nt+{\rm O}(t^2)$ and taking into account that 
$\mu_-+\mu_+=\Delta_n$ and $\mu_-\mu_+=-(\beta_n/2L)^2$, we see from Eqs (\ref{A}) and (\ref{B}) that
$$
A_n=\theta[\mu_-\tau+\ln(1-\mu_-\tau)]+{\rm O}(\tau^2),\qquad
B_n=\frac{(\beta_n/2L)^2\tau}{1-\mu_-\tau}+{\rm O}(\tau^2),
$$
whence
\begin{equation}
S(x,v,\tau)\simeq \sum_{n=0}^\infty\frac{1}{(1-\mu_-\tau)^\theta} 
\exp\left\{-\left[\theta\mu_-+\frac{(\beta_n/2L)^2 v}{1-\mu_-\tau}\right]\tau\right\} 
\cos\left[\left(2n+1\right)\pi x/L\right],\qquad (\tau\ll 1).
\label{fsg_short}
\end{equation}

\begin{figure}
\epsfig{file=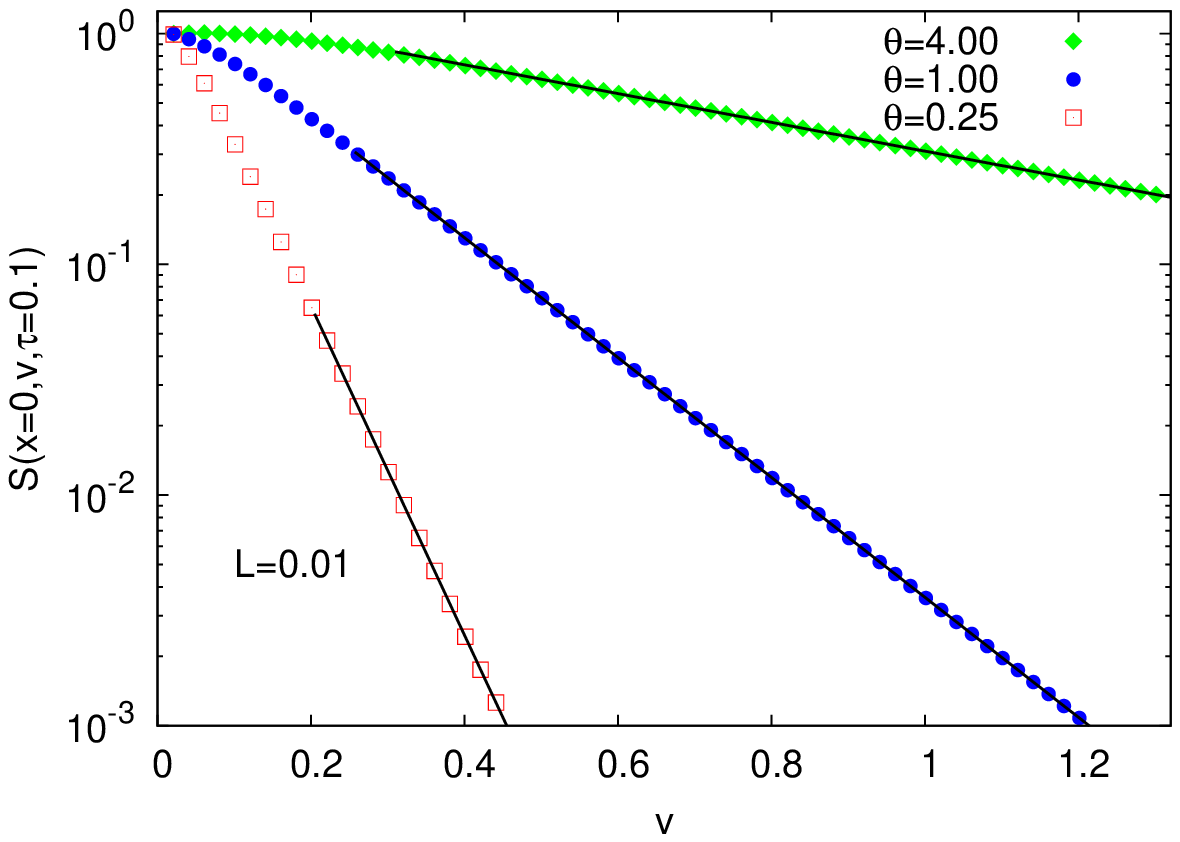,scale=0.6}\epsfig{file=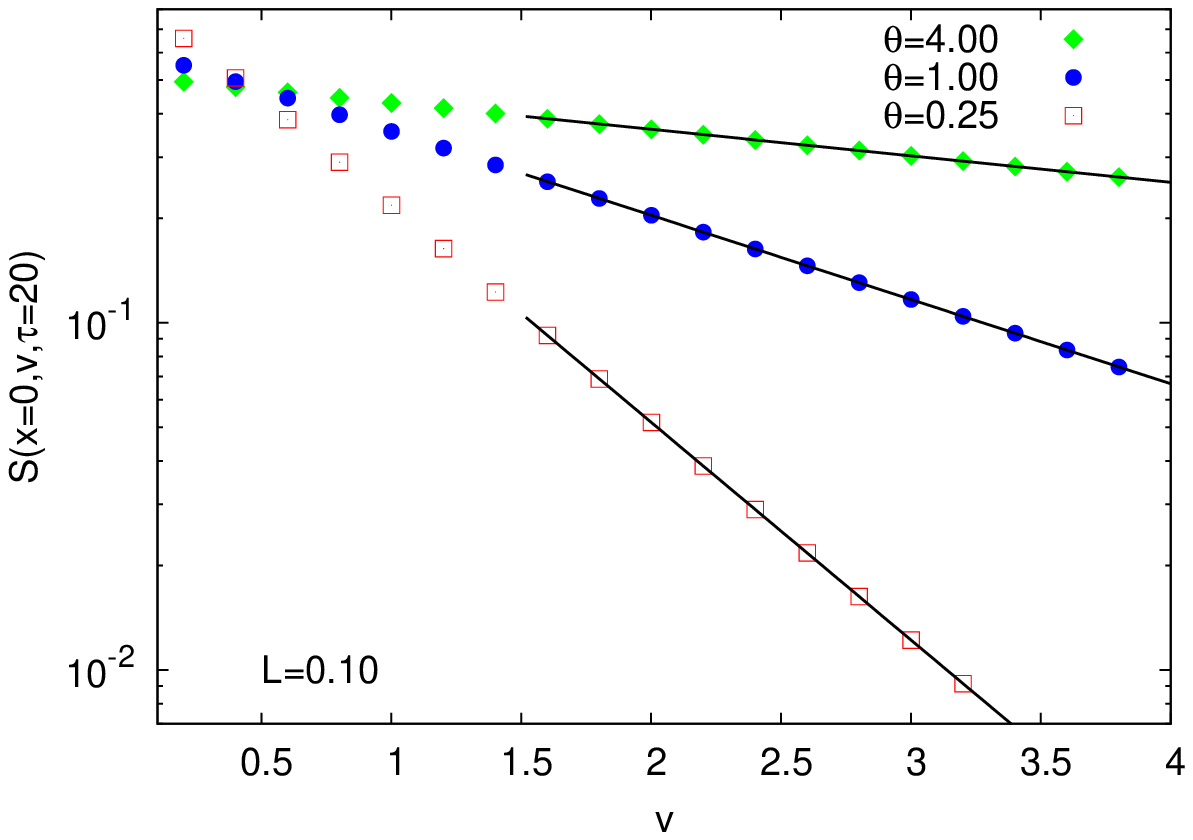,scale=0.6}
\caption{(Color online) The survival probability $S(x,v,\tau)$ given by Eq.~(\ref{complete_solution}) at $x=0$ as a function of the volatility $v$. Left plot shows the case when  $\tau=0.1$ ($t=2.22 \mbox{ days}$) and $L=0.01$. The figure on the right exhibits the case when $\tau=100$ and $L=0.1$. The straight lines correspond to the exponential decay with $v$ mentioned in the main text. Parameters of the model are $\alpha=0.045 \mbox{ day}^{-1}$, $m=0.093 \mbox{ day}^{-1/2}$ and the three different values of the parameter $\theta=(2\alpha/k^2)m^2$ provide three different values for $k$ accordingly.}
\label{spv}
\end{figure}

In Fig.~\ref{spv} we represent the exact SP, Eq.~(\ref{complete_solution}), at $x=0$ and for fixed times, in terms of the volatility. The plots confirm , as hinted by Eqs.~(\ref{fsg_long}) and~(\ref{fsg_short}), that the SP decays exponentially with the volatility for both short and long times. The characteristic exponent of this decay depends on the value of $\theta$, being larger for smaller $\theta$, i.e., larger $k$ (cf. Eq. (\ref{theta})). Moreover, as  $\tau\gg 1$, when the volatility is small, the higher survival probability corresponds to the case when $\theta$ is smaller. This is a distinct behavior with respect to the rest of situations.

\section{The mean escape time}
\label{sec3}

\begin{figure}
\epsfig{file=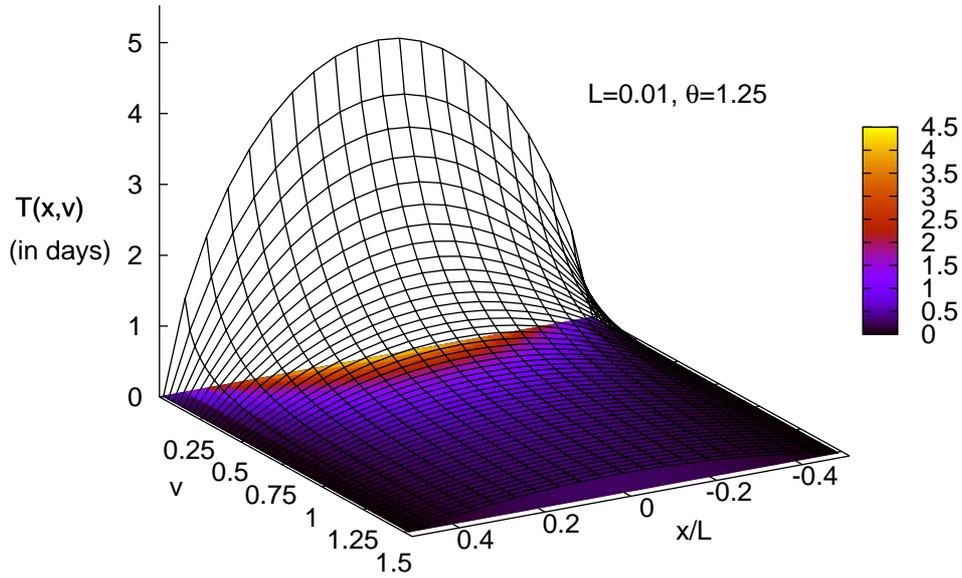}
\caption{(Color online) The mean-escape time $T(x,v)$ given by Eq.~(\ref{2d_met_1}) in terms of return $x$ and volatility variable $v$. Parameters of the model: $\theta=1.25$, $\alpha=0.045 \mbox{ day}^{-1}$, $m=0.093 \mbox{ day}^{-1/2}$ and $k=0.0014 \mbox{ day}^{-1}$.}
\label{met}
\end{figure}

The survival probability $S(x,v,t)$ provides maximal information on the escape problem of the two-dimensional process $(X(t),Y(t))$. Indeed, the probability density function $f(t|x,v)$ of the escape time is related to the SP by \cite{gardiner}
$$
f(t|x,v)=-\frac{\partial S(x,v,t)}{\partial t},
$$
and all moments of the escape time can be obtained through the SP. Thus, for instance, the mean escape (or exit) time (MET) is given by
\begin{equation}
T(x,v)=\int_0^\infty S(x,v,t)dt.
\label{2d_met_def}
\end{equation}

For the Heston model we see from Eq. (\ref{complete_solution}) that the two-dimensional MET, $T(x,v)$, can written in terms of the following Fourier series
\begin{equation}
T(x,v)=\frac{1}{\alpha}\sum_{n=0}^\infty T_n(v) 
\cos\left[\left(2n+1\right)\pi x/L\right],
\label{2d_met_1}
\end{equation}
where
$$
T_n(v)=\gamma_n\int_0^{\infty}\exp\{-A_n(\tau)-B_n(\tau)v\}d\tau.
$$
Using Eqs. (\ref{A})-(\ref{B}) and some simple manipulations, which involve the change of variable $\xi=e^{-\Delta_n\tau}$, yield
\begin{equation}
T_n(v)=\frac{\gamma_n\Delta_n^\theta}{\mu_+^\theta}
\int_0^{1}\frac{\xi^{-1+\mu_-\theta/\Delta_n}}{[1+(\mu_-/\mu_+)\xi]^\theta}
\exp\left\{-\mu_+\left[\frac{1-\xi}{1+(\mu_-/\mu_+)\xi}\right]v\right\}
\label{t_n_v}
\end{equation}

\begin{figure}
\epsfig{file=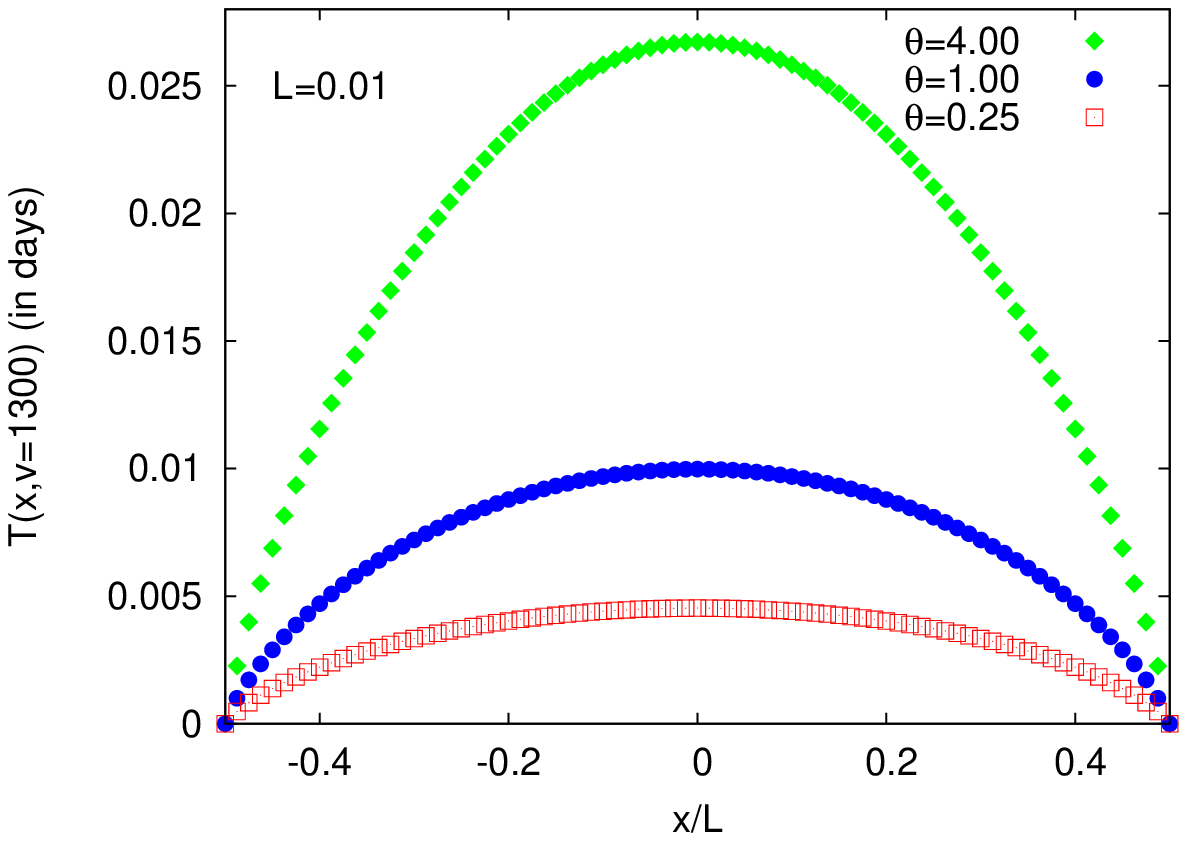,scale=0.6}\epsfig{file=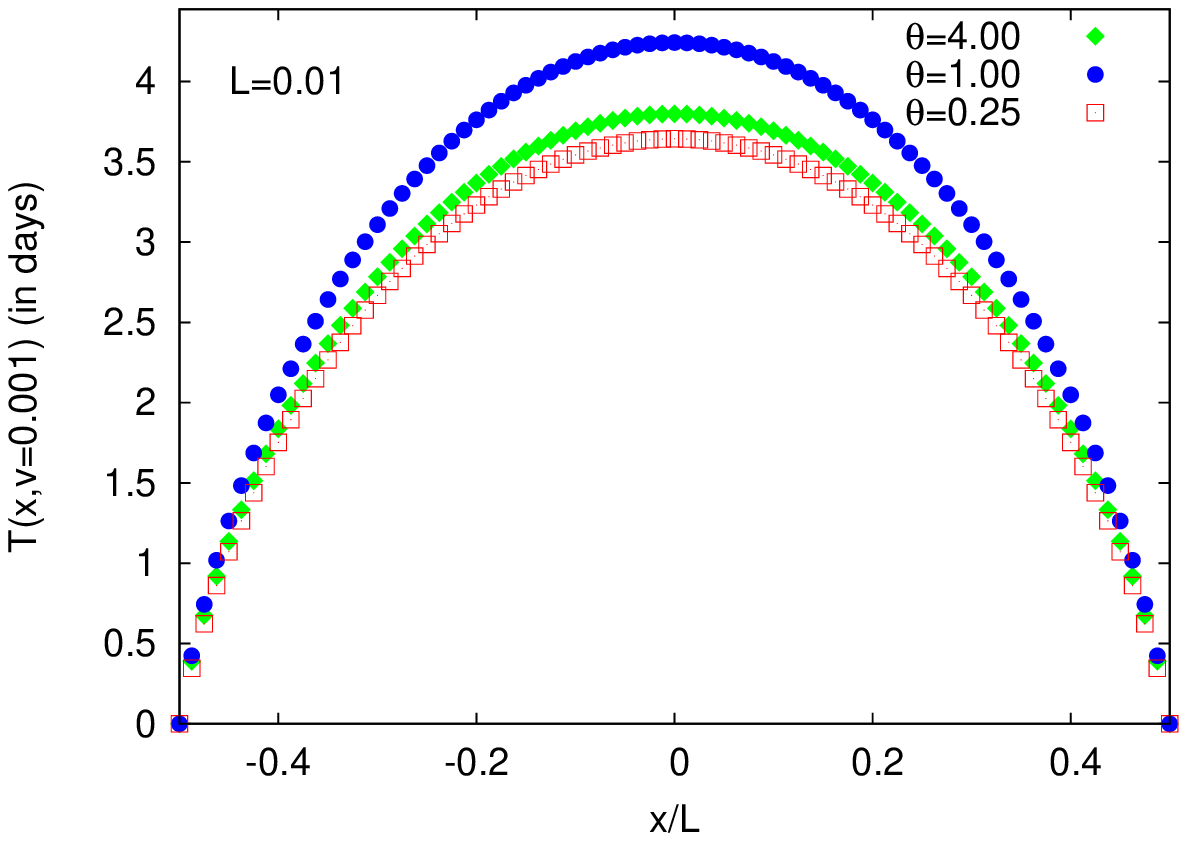,scale=0.6}
\caption{(Color online) The mean-escape time $T(x,v)$ given by Eq.~(\ref{2d_met_1}) as a function of the starting return $x$. Left plot shows the case when $v$ is large showing a perfect hierarchy where larger $\theta$ means larger MET. Right plot shows how the $\theta=1$ case breaks this hierarchical order for small enough values of $v$. Parameters of the model are $\alpha=0.045 \mbox{ day}^{-1}$, $m=0.093 \mbox{ day}^{-1/2}$ and the three different values of the parameter $\theta=(2\alpha/k^2)m^2$ provides three different values for $k$ accordingly.}
\label{met1}
\end{figure}

\begin{figure}
\epsfig{file=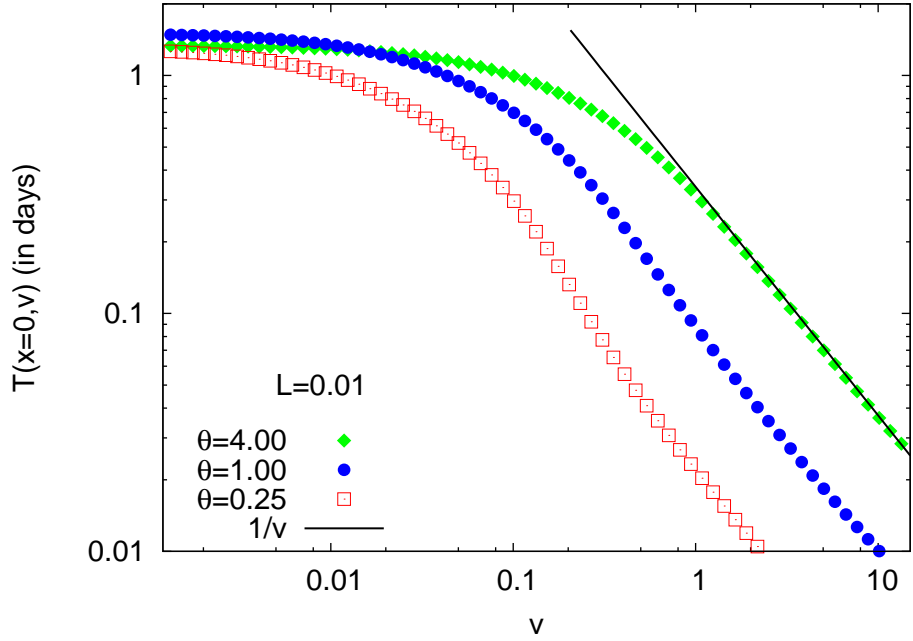}
\caption{(Color online) The mean-escape time $T(x,v)$ given by Eq.~(\ref{2d_met_1}) at $x=0$ as a function of the starting volatility variable $v$. The drawings illustrate that the MET saturates at certain maximum value when $v=0$. On the other hand, the straight line clearly shows that the MET decays as $1/v$ with increasing volatility as we prove in the main text. Parameters of the model are the same than those of Fig. \ref{met1}.}
\label{met2}
\end{figure}

Figure~\ref{met} provides a three-dimensional representation of $T(x,v)$ based on the numerical computation of Eqs.~(\ref{2d_met_1})-(\ref{t_n_v}) \cite{footnote3}. A noticeable aspect worth stressing is  shown in Fig.~\ref{met1} where we two projections of the MET $T(x,v)$ are depicted either for small and large volatility and also for three different values of the normal level $\theta$. Thus when volatility is very large,  $v=1300$, the left plot in Fig.~\ref{met1} shows that the larger $\theta$ corresponds to the longer MET. In the opposite case of very low volatility, $v=0.001$, the right plot shows that this behavior is reversed, for now $\theta=1$ corresponds to a longer $T(x,v)$. This anomaly is also observed in Fig.~\ref{met2} when $v<0.1$.

Having obtained the expression for $T(x,v)$ as given by Eqs. (\ref{2d_met_1}) and (\ref{t_n_v}), let us proceed to elucidate how is the dependence of the MET on the volatility. This is a meaningful question from a practical point of view, for market behavior depends critically on volatility. Intuition tells us that the escape time must tend to zero as the volatility increases and a quick glance both at Eq. (\ref{t_n_v}) and Fig. \ref{met} confirms this, but how is the form of this decrease? On the other hand the behavior of the escape time if the volatility is low is also relevant, will $T(x,v)$ grow without bound as $v\rightarrow 0$? or will it tend to a finite, albeit maximum, value? We will next answer these questions. 

Let us first obtain the behavior of the MET when $v=0$. In such a case Eq. (\ref{t_n_v}) reads
$$
T_n(0)=
\frac{\gamma_n\Delta_n^{\theta-1}}{\mu_+^{\theta}}
\int_0^{1}\frac{\xi^{-1+\mu_-\theta/\Delta_n}}{[1+(\mu_-/\mu_+)\xi]^\theta}d\xi,
$$
and using the following integral representation of the Gauss hypergeometric function \cite{mos}:
\begin{equation}
F(a,b;c;z)=\frac{\Gamma(c)}{\Gamma(b)\Gamma(c-b)}\int_0^1\xi^{b-1}(1-\xi)^{c-b-1}(1-\xi z)^{-a}d\xi \qquad(c>b>0),
\label{gauss}
\end{equation}
we have
\begin{equation}
T_n(0)=\frac{\gamma_n}{\theta\mu_-}\left(\frac{\Delta_n}{\mu_+}\right)^{\theta}
F\left(\theta,\frac{\theta\mu_-}{\Delta_n};1+\frac{\theta\mu_-}{\Delta_n};-\frac{\mu_-}{\mu_+}\right).
\label{t_n_0}
\end{equation}
We therefore see that the mean escape time $T(x,v)$ tends to a finite quantity when $v=0$. 

Let us now turn to the case of increasing volatility. In this situation it is convenient to perform the following change of integration variable in Eq. (\ref{t_n_v}):
$$
z=\frac{1-\xi}{\mu_++\mu_-\xi},
$$
then (recall that $\mu_+\mu_-=(\beta_n/2L)^2$)
$$
T_n(v)=
\gamma_n\int_0^{1/\mu_+}g(z)e^{-(\beta_n/2L)^2v z}dz,
$$
where
$$
g(z)=(1-\mu_+z)^{-1+\mu_-\theta/\Delta_n}(1+\mu_-z)^{-1+\theta-\mu_-\theta/\Delta_n}.
$$
As $v\rightarrow\infty$ the exponential term falls off quickly and we may safely change the upper integration limit $1/\mu_+$ by $\infty$. Using then Watson's lemma we write \cite{erdelyi}
\begin{equation}
T_n(v)\sim \gamma_n\sum_{k=o}^\infty g^{(k)}(0)\frac{(2L/\beta_n)^{2k}}{v^{k+1}}.
\label{t_n_infty}
\end{equation}

Up to the leading order ($g(0)=1$)
$$
T_n(v)\sim \gamma_n\left(2L/\beta_n\right)^2(1/v)+{\rm O}\left(1/v^2\right),
$$
or (cf. Eqs. (\ref{gamma}) and (\ref{theta}))
\begin{equation}
T_n(v)\sim \frac{16\alpha^2L^2}{\pi^3 k^2}\frac{(-1)^n}{(2n+1)^3}(1/v)+{\rm O}\left(1/v^2\right).
\label{t_n_infty_1}
\end{equation}
Therefore
$$
T(x,v)\sim \frac{16\alpha L^2}{\pi^3k^2}(1/v)\sum_{n=0}^{\infty}\frac{(-1)^n}{(2n+1)^3}\cos[(2n+1)\pi x/L]+
{\rm O}\left(1/v^2\right).
$$
The series on the right can be summed with the result \cite{footnote4}
\begin{equation}
T(x,v)\sim\frac{2\alpha}{k^2v}[(L/2)^2-x^2]+O(1/v^2).
\label{t_infty}
\end{equation}
This is a remarkable result since shows that for large volatility the MET has the same form than that of the Wiener process (see Sect. \ref{sec4}). Moreover the two-dimensional MET decreases linearly as $1/v$.

The behavior of $T(x,v)$ with volatility is clearly seen from the numerical evaluation of the exact MET given by Eqs. (\ref{2d_met_1}) and (\ref{t_n_v}). Figure~\ref{met2} shows, in a log-log scale, how the MET saturates to a maximum when $v$ tends to zero while for large volatility $T(x,v)$ is well  fitted with a power law with exponent -1 which confirms the asymptotic expression (\ref{t_infty}).

\section{Averaging the volatility}
\label{sec4}

In real financial data the volatility is, in fact, a hidden variable which has to be measured in an indirect way \cite{zoltan}. It is therefore of great significance to know whether the price of an asset remains inside a given interval regardless its volatility. In physics the analog to this question would be knowing the survival probability for the position of a Brownian particle without worrying about its velocity \cite{mpo_pre}. Considering the entanglement between return and volatility (or position and velocity) this is certainly a difficult question and one often has to rely on approximate answers. Fortunately the latter is not the case in the Heston model as we shall see next.

\subsection{The survival probability of the return}
\label{subsec4a}

\begin{figure}
\epsfig{file=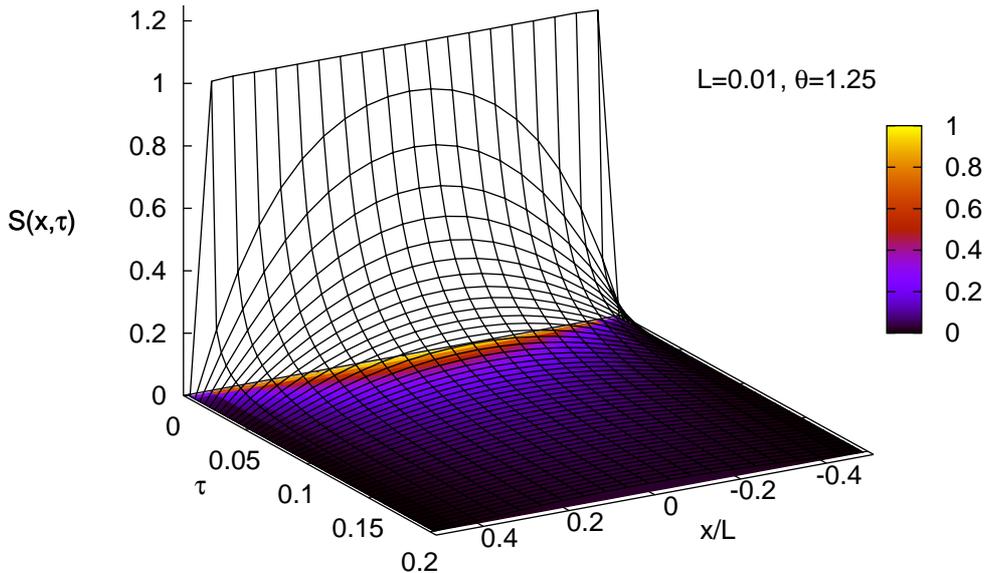}
\caption{(Color online) The survival probability of the return, $S(x,\tau)$, given by Eq.~(\ref{exact_return}) with $L=0.01$ as a function of the return $x$ and time $\tau=\alpha t$. Parameters of the model: $\theta=1.25$, $\alpha=0.045 \mbox{ day}^{-1}$, $m=0.093 \mbox{ day}^{-1/2}$ and $k=0.0014 \mbox{ day}^{-1}$.}
\label{sx}
\end{figure}

In order to obtain the SP of the return, $S(x,\tau)$, regardless the value of the volatility, we have to average the volatility away from  $S(x,v,\tau)$. We will do this by assuming that, at the time we measure the return, the volatility process has reached the stationary state \cite{footnote_stat}. We therefore define $S(x,\tau)$ as the average:
\begin{equation}
S(x,\tau)=\int_0^\infty S(x,v,\tau)p_{\rm st}(v)dv,
\label{return_SP}
\end{equation}
where $p_{\rm st}(v)$ is the stationary probability density of the volatility. For the Hestson model this density is
the normalized solution of the Fokker-Planck equation
$$
\frac{d}{dv}\left[(v-\theta)+\frac{d}{dv}v\right]p_{\rm st}(v)=0,
$$
which is given by the Gamma distribution:
\begin{equation}
p_{\rm st}(v)=\frac{1}{\Gamma(\theta)}v^{\theta-1}e^{-v}.
\label{stat}
\end{equation}
Note that $\theta$ is the stationary variance of the volatility variable $v$. Indeed, from Eq. (\ref{stat}) we see at once that $\theta=\langle v^2\rangle_{\rm st}-\langle v\rangle_{\rm st}^2$. Observe also the changing shape of the stationary distribution (specially as $v\rightarrow 0$) according to whether $\theta<1$ or $\theta>1$; a fact that, as we shall see below, has consequences on the behavior of the MET. 

From Eqs. (\ref{fss}) and (\ref{return_SP}) we get
\begin{equation}
S(x,\tau)=\sum_{n=0}^\infty S_n(\tau)\cos\left[\left(2n+1\right)\pi x/L\right],
\label{f_r}
\end{equation}
where
$$
S_n(\tau)=\int_0^\infty S_n(v,\tau)p_{\rm st}(v)dv,
$$
which, after making use of Eqs. (\ref{solution_1}) and (\ref{stat}), yields
\begin{equation}
S_n(\tau)=\gamma_n\frac{e^{-A_n(\tau)}}{[1+B_n(\tau)]^\theta}.
\label{sn_1}
\end{equation}

We will write this Fourier coefficient in a more convenient form. Let us first note that by applying Eq. (\ref{A}) we can write
\begin{equation}
e^{-A_n(\tau)}=\left(\frac{\Delta_n e^{-\mu_-\tau}}{\mu_++\mu_-e^{-\Delta_n\tau}}\right)^\theta.
\label{expA}
\end{equation}
On the other hand, from Eq. (\ref{B}) we see
$$
1+B_n=\frac{\mu_+(1+\mu_-)+\mu_-(1-\mu_+)e^{-\Delta_n\tau}}{\mu_++\mu_-e^{-\Delta_n\tau}};
$$
but $1+\mu_-=\mu_+$ and $1-\mu_+=-\mu_-$ (cf. Eq. (\ref{delta})). Hence 
\begin{equation}
1+B_n(\tau)=\frac{\mu_+^2-\mu_-^2e^{-\Delta_n\tau}}{\mu_++\mu_-e^{-\Delta_n\tau}}.
\label{1+B}
\end{equation}
Plugging Eqs. (\ref{expA})-(\ref{1+B}) into Eq. (\ref{sn_1}) we have
\begin{equation}
S_n(\tau)=\gamma_n\left(\frac{\Delta_n e^{-\mu_-\tau}}{\mu_+^2-\mu_-^2e^{-\Delta_n\tau}}\right)^\theta,
\label{sn}
\end{equation}
and therefore
\begin{equation}
S(x,\tau)=\sum_{n=0}^\infty \gamma_n
\left(\frac{\Delta_n e^{-\mu_-\tau}}{\mu_+^2-\mu_-^2e^{-\Delta_n\tau}}\right)^\theta
\cos\left[\left(2n+1\right)\pi x/L\right],
\label{exact_return}
\end{equation}
which constitutes the exact expression for the SP of the return.

We will now show the asymptotic time behavior of $S(x,\tau)$. We easily see from Eq. (\ref{exact_return}) that for long times, $\tau\gg 1$, the asymptotic form of the SP is
\begin{equation}
S(x,\tau)\simeq \sum_{n=0}^\infty \gamma_n
\left(\frac{\Delta_n}{\mu_+^2}\right)^\theta e^{-\theta\mu_-\tau}
\cos\left[\left(2n+1\right)\pi x/L\right], \qquad (\tau\gg 1),
\label{exact_return_long}
\end{equation}
while for short times $\tau\ll 1$ and after taking into account (cf. Eq. (\ref{delta})) 
$$
\mu_+^2-\mu_-^2e^{-\Delta_n\tau}=\Delta_n(1+\mu_-^2\tau)+{\rm O}(\tau^2),
$$
we get
\begin{equation}
S(x,\tau)\simeq \sum_{n=0}^\infty \gamma_n
\frac{e^{-\theta\mu_-\tau}}{(1+\mu_-^2\tau)^\theta}
\cos\left[\left(2n+1\right)\pi x/L\right], \qquad (\tau\ll 1).
\label{exact_return_short}
\end{equation}

\begin{figure}
\epsfig{file=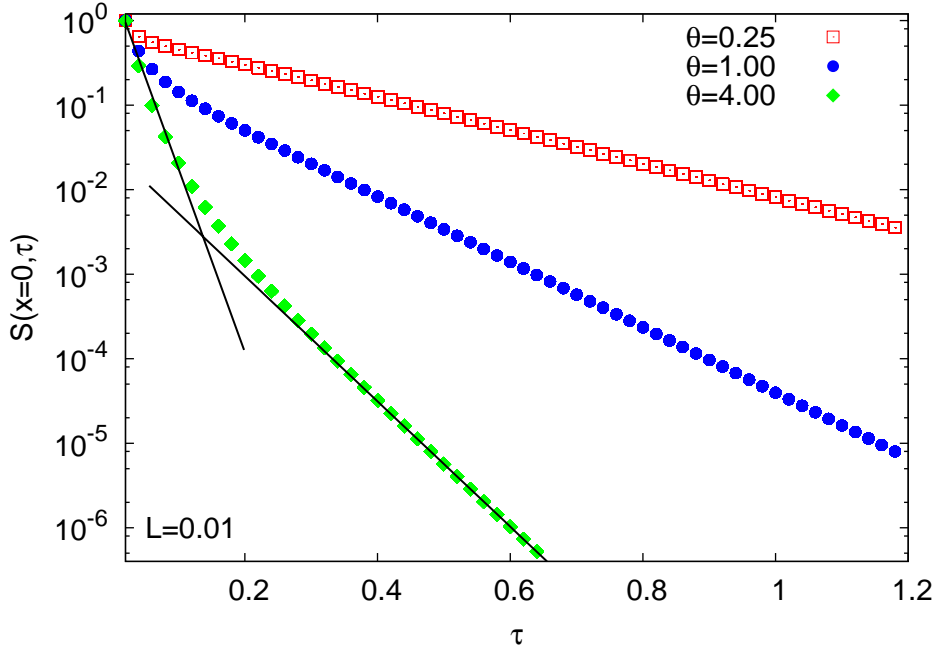}
\caption{(Color online) The survival probability $S(x,\tau)$ given by Eq.~(\ref{exact_return}) with $L=0.01$ as a function of the time $\tau=\alpha t$ and starting from the return midpoint $x=0$. Parameters of the model are: 
$\alpha=0.045 \mbox{ day}^{-1}$ and $m=0.093 \mbox{ day}^{-1/2}$. Notice that since 
$\alpha=0.045 \mbox{ day}^{-1}$ then $\tau=1$ corresponds to an actual time of $t\simeq 22\mbox{ days}$. The straight lines plotted on the lowest curve are the asymptotic approximations given by 
Eqs. (\ref{exact_return_long}) ($\tau\gg 1$) and (\ref{exact_return_short}) ($\tau\ll 1$).}
\label{sx1}
\end{figure}

The approximate expressions for $S(x,\tau)$ given in Eqs. (\ref{exact_return_long})-(\ref{exact_return_short}) suggest an exponential decay (essentially governed by the normal level $\theta$) either for short and long times. This is confirmed by the numerical evaluation of the exact SP, 
Eq. (\ref{exact_return}), which we present in Fig. ~\ref{sx1}. We clearly see there two different exponential decays which match those shown in 
Eqs. (\ref{exact_return_long})-(\ref{exact_return_short}).

\subsection{The mean escape time of the return}
\label{subsec4b}

\begin{figure}
\epsfig{file=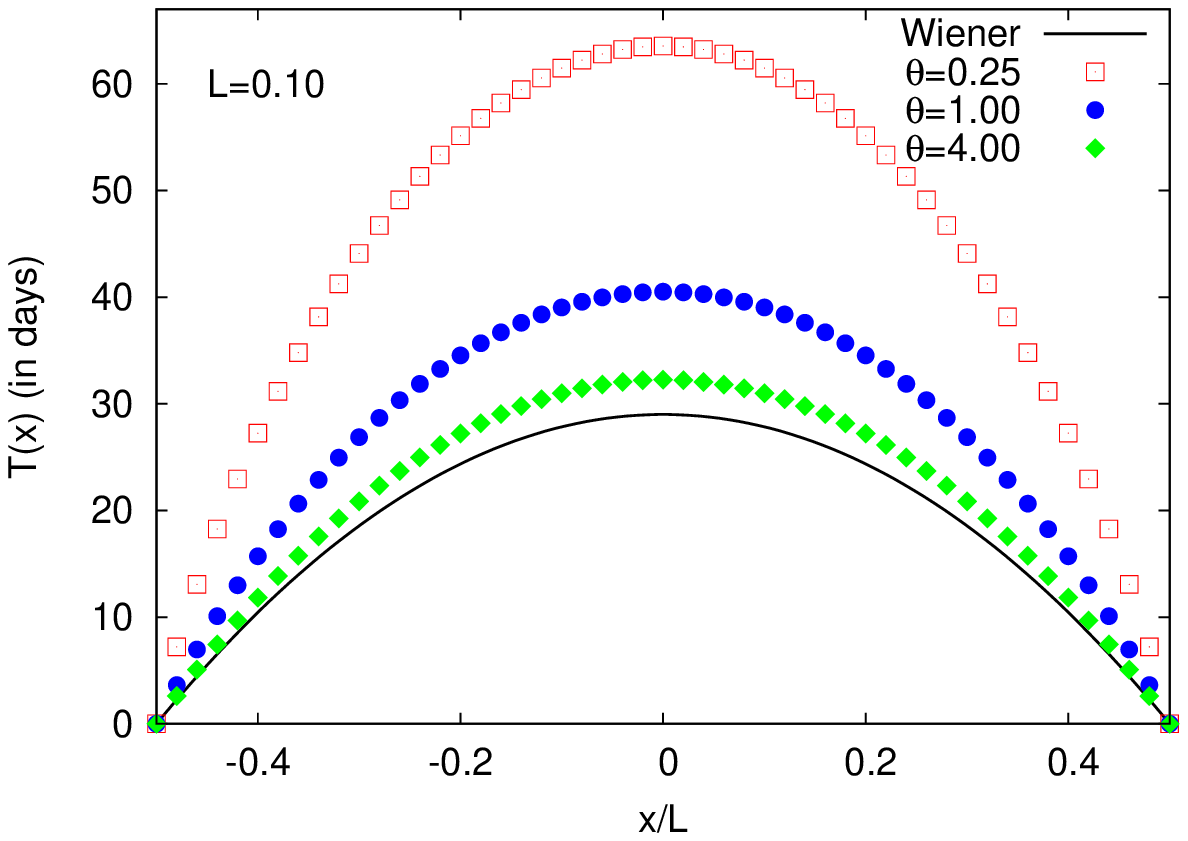,scale=0.6}\epsfig{file=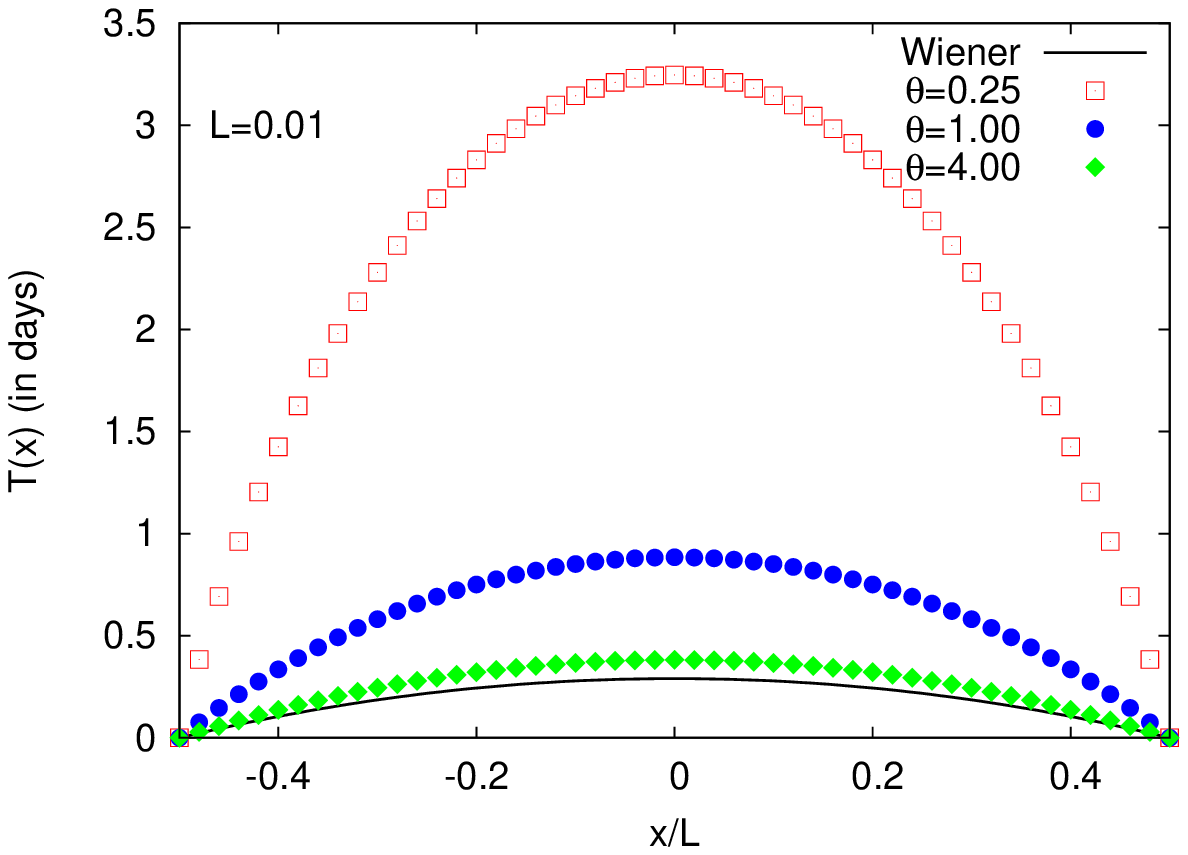,scale=0.6}
\caption{(Color online) The mean-escape time $T(x)$ based on the exact expression (\ref{exact_met}) as a function of the starting return $x$. The left figure shows the case when $L=0.1$ while the right plot shows the case when $L=0.01$. In both cases the larger the $\theta$ the lower the MET. We also draw the MET corresponding to the Wiener process (note that the latter is always shorter than Heston's MET). Parameters of the model are the same than those of Fig. \ref{met1}.}
\label{metv}
\end{figure}

In terms of the survival probability $S(x,t)$ the mean escape time is given by
$$
T(x)=\int_0^\infty S(x,t)dt.
$$
Combining this equation with Eq. (\ref{exact_return}) we see that $T(x)$ is written as a Fourier series of the form
\begin{equation}
T(x)=\frac{1}{\alpha}\sum_{n=0}^\infty T_n 
\cos\left[\left(2n+1\right)\pi x/L\right],
\label{exact_met_1}
\end{equation}
with Fourier coefficients given by
\begin{equation}
T_n=\gamma_n\Delta_n^\theta\int_0^{\infty}
\left(\frac{e^{-\mu_-\tau}}{\mu_+^2-\mu_-^2e^{-\Delta_n\tau}}\right)^\theta d\tau.
\label{t_n_1}
\end{equation}
The integral appearing in the right hand side of this equation is evaluated by performing the change of variables $\xi=e^{-\Delta_n\tau}$. We have
$$
T_n=\frac{\gamma_n\Delta_n^{\theta-1}}{\mu_+^{2\theta}}
\int_0^{1} \xi^{-1+\theta\mu_-/\Delta_n}\left[1-\left(\mu_-/\mu_+\right)^2\xi\right]^{-\theta}d\xi,
$$
and taking into account the integral representation of the Gauss hypergeometric function given in Eq. (\ref{gauss}) we get
$$
T_n=\frac{\gamma_n}{\theta\mu_-}\left(\frac{\Delta_n}{\mu_+^2}\right)^\theta
F\left(\theta,\frac{\theta\mu_-}{\Delta_n};1+\frac{\theta\mu_-}{\Delta_n}; \frac{\mu_-^2}{\mu_+^2}\right).
$$
Finally, the MET is given by (see Fig. \ref{metv})
\begin{equation}
T(x)=\frac{1}{\alpha\theta}
\sum_{n=0}^\infty\frac{\gamma_n}{\mu_-}\left(\frac{\Delta_n}{\mu_+^2}\right)^\theta
F\left(\theta,\frac{\theta\mu_-}{\Delta_n};1+\frac{\theta\mu_-}{\Delta_n}; \frac{\mu_-^2}{\mu_+^2}\right) 
\cos\left[\left(2n+1\right)\pi x/L\right].
\label{exact_met}
\end{equation}

From a practical point of view an interesting property to look at is the behavior of the MET as a function of the span $L$ specially for short and large values of $L$, being the latter closely related to financial defaults or uprisings  depending on the sign of $x$. We will thus consider the two limiting cases: (a) $L\rightarrow 0$ and (b) $L\rightarrow \infty$.

\begin{figure}
\epsfig{file=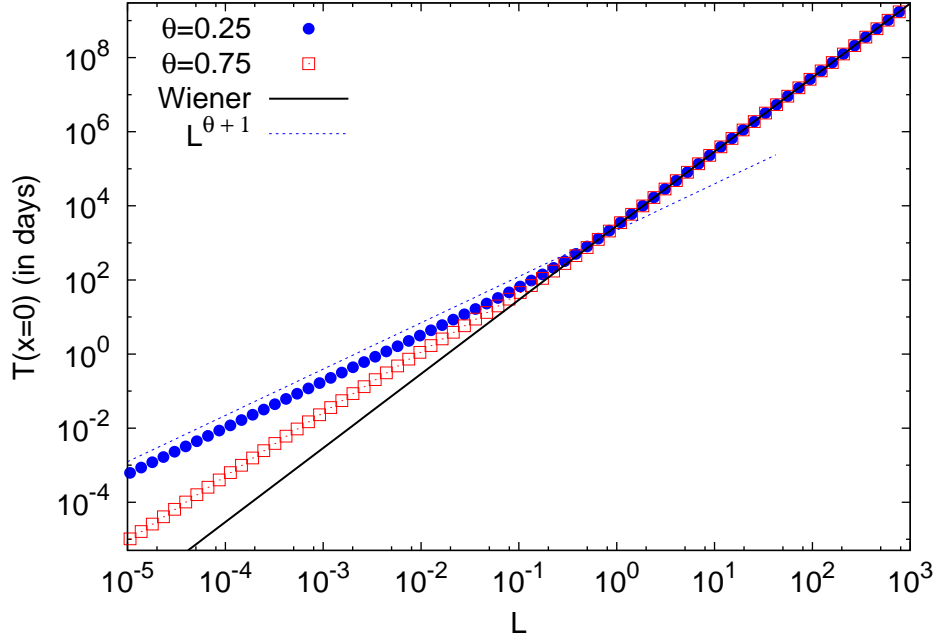}
\caption{(Color online) The mean-escape time $T(x=0)$ given by Eq.~(\ref{exact_met}) as a function of the span $L$ when $\theta<1$. The solid lines corresponds to $L^2$. Parameters of the model are the same than those of Fig. \ref{met1}}
\label{metL}
\end{figure}

\begin{figure}
\epsfig{file=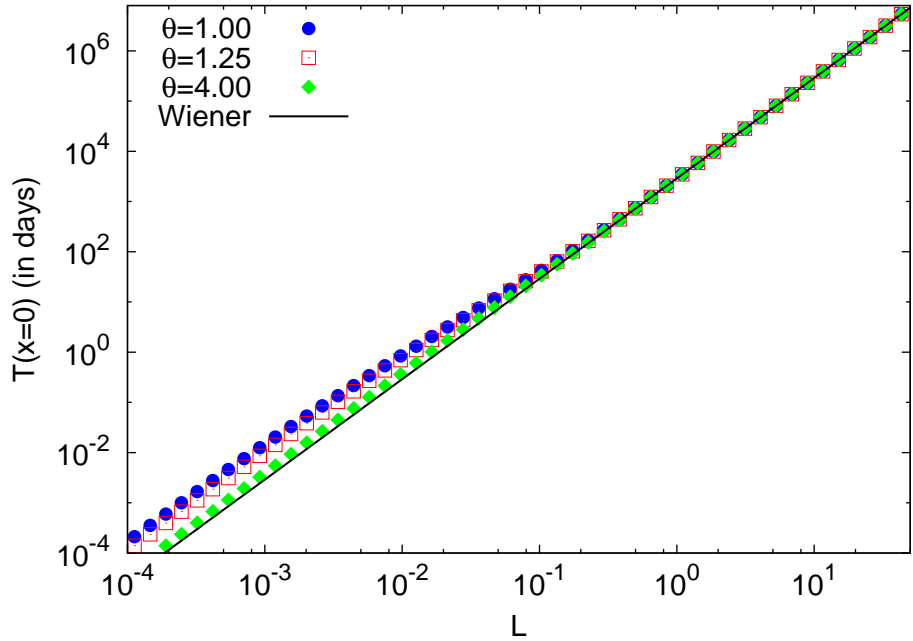}
\caption{(Color online) The mean-escape time $T(x=0)$ given by Eq.~(\ref{exact_met}) as a function of the span $L$ when $\theta\geq 1$. Parameters of the model are the same than those of Fig. \ref{met1}.}
\label{metL1}
\end{figure}

(a) In the case of small span the Taylor expansion as $L\rightarrow 0$ of Eq. (\ref{exact_met}) leads to the following asymptotic expression (see Appendix \ref{appB} for details)
\begin{equation}
T(x)\sim \begin{cases} 
L^{\theta+1}  & \theta<1, \\
-L^2\ln L & \theta=1, \qquad(L\rightarrow 0) \\
L^2  & \theta>1.
\end{cases}
\label{TL0}
\end{equation}  
We see that in this case the behavior of the MET is governed by the (dimensionless) normal level $\theta$ which coincides with the stationary variance of the volatility variable $v$. Let us recall that a similar situation arises for the stationary distribution since, as seen in Eq. (\ref{stat}), $p_{\rm st}(v)$ behaves in a different way  according to whether the normal level is greater or lower than $1$. Note that (cf. Eq. (\ref{theta})) $\theta<1$ implies $m^2<k^2/\alpha$, that is, volatility fluctuations --represented by the vol-of-vol $k$-- are wilder than the tendency toward the normal level given by $\alpha m^2$. On the other hand, when this tendency is greater than the volatility fluctuations ($\theta>1$) the MET grows quadratically with $L$ independent of the normal level $m$ but with slope depending on the vol-of-vol through the combination 
$k^2/\alpha$. All of this is exemplified in Figs. \ref{metL}-\ref{metL1} where we plot, based on the exact expression (\ref{exact_met}), the MET as a function of the span $L$. 

(b) Let us now look at the behavior of the MET with increasing span. Unfortunately this case is more  difficult to deal with since $L$ appears in the Fourier series solution basically through the combination $(2n+1)/L$ and any effect due to $L\rightarrow\infty$ is neutralized by increasing values of $n$ which, in turn, are needed to sum the Fourier series. The case (a) above turns out to be workable because the limits $L\rightarrow 0$ and $n\rightarrow\infty$ are compatible.

On the other hand, numerical calculations of the exact MET (\ref{exact_met}) shown in Figs. \ref{metL}-\ref{metL1}  clearly indicate that the MET grows quadratically with the span regardless the value of the normal level $m$:
\begin{equation}
T(x)\sim L^2, \qquad (L\rightarrow\infty).
\label{large_span}
\end{equation}
We remind that we have already encountered this behavior at the end of Sect. \ref{sec3} when analyzing the two-dimensional MET, $T(x,v)$, for large volatility (cf. Eq. (\ref{t_infty})). In the Appendix \ref{appC} we justify this quadratic growth by means of a heuristic argument.

We also note that now, contrary to the case of small span, the slope is independent of $k^2/\alpha$ and all the cases which have the same $m$ merge into a single curve (cf. Figs. \ref{metL}-\ref{metL1})

\subsection{The Wiener process}
\label{subsec4c}

For many years the most ubiquitous market model has been the geometric Brownian motion which was proposed by Osborne in 1959 \cite{cootner}. In this model the price $P(t)$ obeys the stochastic differential equation
$$
\frac{dP(t)}{P(t)}=\nu dt+\sigma dW(t),
$$
where $\nu$ is a constant drift, $\sigma$ is the volatility (a constant as well) and $W(t)$ is the Wiener process. In terms of the zero-mean return $X(t)$ defined in Eq. (\ref{zero-mean}), the model reads
$$
dX(t)=\sigma dW(t).
$$
In other words, $X(t)$ is the Wiener process with variance $\sigma^2$. 

In view of the widespread use of this market model among practitioners and even academicians \cite{lo}, we find it convenient to compare the findings on the escape problem of the return discussed in this section with those of the Wiener process. This, in turn, may provide a test on the appropriateness of the assumption of stochastic volatility for real market models. 

Let us thus suppose that the zero-mean return is described by the Wiener process and denote by $S_0(x,t)$ its survival probability inside the interval $-L/2\leq X(t)\leq L/2$. This function obeys the equation \cite{gardiner}
$$
\frac{\partial S_0}{\partial t}=\frac{1}{2}\sigma^2 \frac{\partial^2 S_0}{\partial x^2},
$$
with initial and boundary conditions
$$
S_0(x,0)=1,\qquad S_0(\pm L/2,t)=0.
$$
Proceeding as we have done before we look for a solution to this problem in terms of a Fourier series. In this way one easily obtains
\begin{equation}
S_0(x,t)=
\frac{4}{\pi}\sum_{n=0}^\infty\frac{(-1)^n}{2n+1}\exp\left\{-[\pi L\sigma(2n+1)]^2t/2\right\}\cos\left[\left(2n+1\right)\pi x/L\right],
\label{surv_wiener}
\end{equation}
and the MET is
\begin{equation}
T_0(x)=\frac{1}{\sigma^2}[(L/2)^2-x^2].
\label{met_wiener}
\end{equation}

\begin{figure}
\epsfig{file=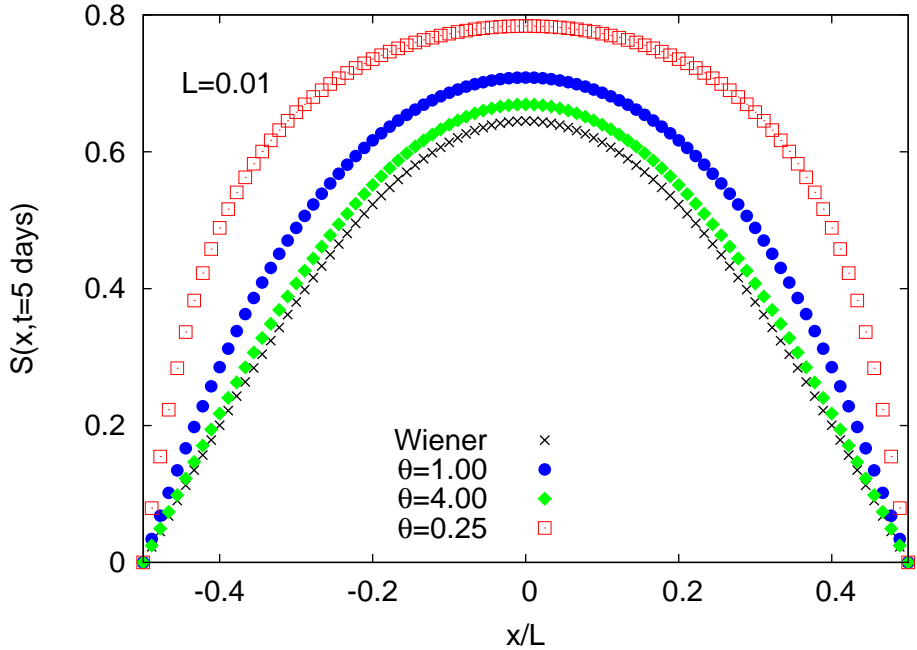}
\caption{(Color online) Return survival probability as a function of the scaled starting return $x/L$ when $L=0.05$. We represent the $S(x,\tau/\alpha=5\mbox{ days})$ of the Heston model given by Eq.~(\ref{return_SP}) for several values of $\theta$. We also plot the SP corresponding to the Wiener process, $S_0(0,t=5 \mbox{ days})$, given in Eq.~(\ref{surv_wiener}). Parameters of the Heston model are $\alpha=0.045 \mbox{ day}^{-1}$ and $m=0.093 \mbox{ day}^{-1/2}$. For the Wiener case we suppose that volatility is equal to the normal level $\sigma=m$.}
\label{sv-wiener-x}
\end{figure}

\begin{figure}
\epsfig{file=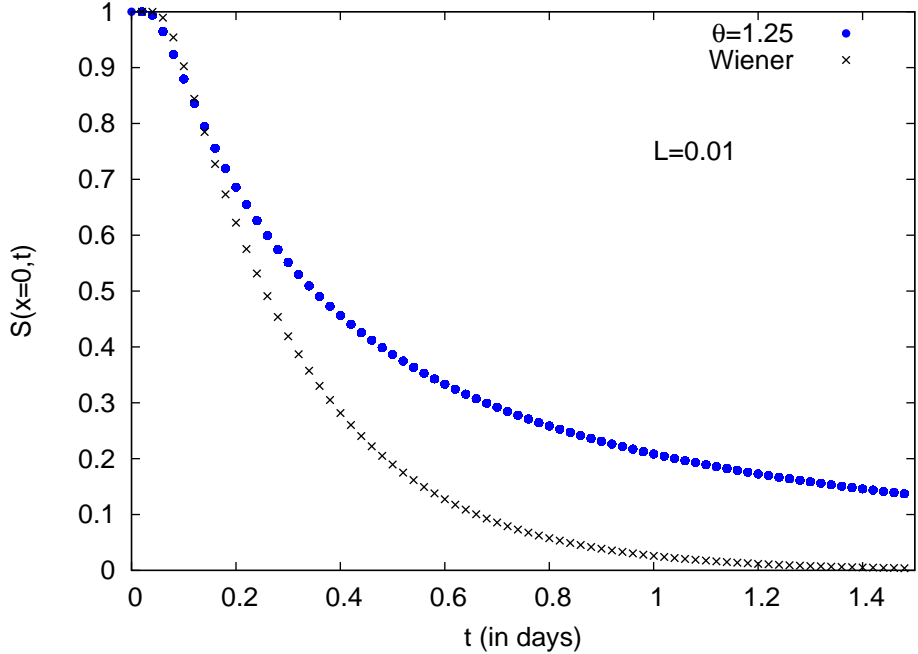}
\caption{(Color online) Return survival probability as a function of the scaled time $\tau$. We represent the decay $S(0,\alpha t)$ with time for the Heston model given in Eq.~(\ref{return_SP}) when $\theta=1.25$ in comparison with the Wiener model SP, $S_0(0,t)$, provided by Eq.~(\ref{surv_wiener}). In both cases we assume that starting return is the midpoint of the interval $L=0.01$. Parameters of the model are the same than those of Fig. \ref{met1}.}
\label{sv-wiener-t}
\end{figure}

In Fig. \ref{sv-wiener-x} we plot the $S_0(x,t)$ given by Eq. (\ref{surv_wiener}) in terms of the return $x$ and for a fixed time $t=5$ days. In Fig. \ref{sv-wiener-t} we do the same but as a function of time and for a fixed return $x=0$. In the both figures we also represent the Heston's SP, $S(x,t)$, given in Eq.~(\ref{exact_return}). We see that the survival probability is always higher under stochastic volatility than when the volatility is constant; although for a greater normal level $\theta$ this difference becomes smaller. 

Thus, for instance (cf. Fig. \ref{sv-wiener-t}) when $\theta=1.25$ the survival probability of the Wiener process in one day starting at $x=0$ with span $L=0.01$ is just $S_0(0, 1 \mbox{ day})=0.026$; for the Heston model this probability is $8$ times higher: $S(0, 1 \mbox{ day})=0.208$. In two days the difference is even higher: $S_0(0, 2 \mbox{ days})=0.0005$ versus $S(0, 2 \mbox{ days})=0.095$. 

This difference is also detected in the MET. Thus in Fig. \ref{metv} we see that the Heston's MET is invariably longer than that of Wiener's. In other words, the Wiener process escapes faster than the Heston SV model. 

Therefore, and contrary to intuition, the assumption of stochastic volatility, notwithstanding occasional bursts, seems to stabilize prices after a certain number of time steps.

\section{Summary and conclusions}
\label{sec5}

We have studied the escape problem of the return under the assumption of a stochastic volatility given by the Heston model. The problem is fully characterized by knowing the survival probability, $S(x,y,t)$, of the bidimensional process $(X(t),Y(t))$ inside the strip:
$$
-L/2\leq X(t)\leq L/2,\qquad 0<Y(t)<\infty,
$$
where $X(t)$ is the zero-mean return and $Y(t)$ is the volatility variable which, for the Heston model, is related to the volatility $\sigma$ by $\sigma(t)=\sqrt{Y(t)}$. The survival probability obeys the backward Fokker-Planck equation (\ref{fpe}) with initial and boundary conditions given in Eq. (\ref{ibc}). We have been able to exactly solve the problem by means of the Fourier series expansion given in Eq. (\ref{complete_solution}).

Once we have the solution for $S(x,y,t)$ another interesting and most useful quantity to know is the mean escape time. For the entire process given by the return and the volatility, the MET $T(x,y)$ is also exactly given by a Fourier series (cf. Eqs. (\ref{2d_met_1})-(\ref{t_n_v})). We have shown that, as the volatility decreases, $T(x,y)$ tends toward a maximum, albeit finite, value. Moreover, as the volatility increases the MET decreases following the hyperbola $1/y$ which is quite remarkable because this is exactly the behavior of the MET with the volatility had the retun followed the Wiener process (i.e., constant volatility) instead of the Heston model. 

Real financial data consist of time series of prices and volatility is not directly recorded and only observed in an indirect way. This hidden character makes it worth averaging out volatility from the expressions of $S(x,y,t)$ and $T(x,y)$ and thus solving the escape problem for the return alone. The assumption to be made is that the volatility has reached the stationary state; in the Heston model the latter is characterized by the Gamma distribution, Eq. (\ref{stat}). 

Following this way we have obtained exact expressions of $S(x,t)$, Eq. (\ref{exact_return}), and  $T(x)$, Eq. (\ref{exact_met}), both in terms of Fourier series. The SP has two different exponential decays: one for long times and another, which is faster, for short times. We have been able to get analytical expressions of both decays.

We have next analyzed the behavior of $T(x)$ as a function of the span $L$, specially for short and large values of $L$. The latter case is specially significant because large values of $L$ are associated to financial uprisings or defaults. We have shown that the behavior of the MET as $L\rightarrow 0$ depends on the normal level $\theta$ and it is given in Eq. (\ref{TL0}). On the other hand, when $L\rightarrow\infty$ the MET grows as $L^2$ independently of the normal level. 

Therefore, when $\theta<1$ (i.e., if volatility fluctuations are greater than the tendency toward the normal level) we have a ``crossover'' in the MET, from $L\rightarrow 0$ to $L\rightarrow\infty$, of the form:
$$
T(x)\sim L^{\theta+1}\longrightarrow T(x)\sim L^2, \qquad(\theta<1).
$$
On the other hand for $\theta>1$ (the tendency to relax toward the normal level is now stronger than the fluctuations of the volatility) there is no such crossover, since $T(x)\sim L^2$ for both small and large values of the span. Again, this quadratic dependence is the same as if the return would had been described by the ordinary Wiener process. 

We have finally compared the return SP and MET to those provided by the Wiener process. In other words, we have confronted the escape problem of the Heston model with that of the Wiener process. Our main finding is that Heston's SP is bigger and Heston's MET is longer than those corresponding to the Wiener process. This, at first sight, is contrary to intuition because a random volatility, despite occasional bursts, would seem to stabilize prices to a greater extend than a constant volatility. However, let us recall that in $S(x,t)$ and $T(x)$ the volatility has been averaged around its mean value that is precisely the normal level $\theta$ and, if $\theta$ is not very large, this fact may be the responsible for the stabilization of the return. 

A final remark. The Heston model is one among several possible candidates aimed to describe a realistic price dynamics. The question of which SV model is more appropriate as a market model is still an open question \cite{qf}. We have chosen the Heston model to carry on the present development because, as we have seen, it allows for an exact treatment. In a forthcoming work we will present an approximation scheme in order to study the escape problem for a wider class of models.

\acknowledgments 
Partial financial support from Direcci\'on General de Investigaci\'on under contract No. FIS2006-05204 is acknowledged.

\appendix

\section{Functions $A_n(\tau)$ and $B_n(\tau)$}
\label{appA}

To obtain the functions $A_n(\tau)$ and $B_n(\tau)$ we must solve the Riccatti equation 
\begin{equation}
\dot{B_n}=-B_n-B_n^2+(\beta_n/2L)^2,
\label{a1}
\end{equation}
with initial condition $B_n(0)=0$. To this end we define a new function $Z_n(\tau)$ related to $B_n(\tau)$ by
$$
B_n(\tau)=\frac{\dot{Z}}{Z}.
$$
then $Z(\tau)$ obeys the linear equation
$$
\ddot{Z}+\dot{Z}-(\beta_n/2L)^2 Z=0,
$$
whose solution reads
$$
Z(\tau)=C_1e^{\mu_-\tau}+C_2e^{-\mu_+\tau},
$$
where $C_1$ and $C_2$ are arbitrary constants and
$$
\mu_{\pm}=(\Delta_n\pm 1)/2, \qquad \Delta_n=\sqrt{1+4\beta_n^2}.
$$
The expression for $B_n(\tau)$ is thus given by
$$
B_n(\tau)=\frac{\mu_-e^{\mu_-\tau}-(C_2/C_1)\mu_+e^{-\mu_+\tau}}{e^{\mu_-\tau}+(C_2/C_1)e^{-\mu_+\tau}},
$$
and the initial condition $B_n(0)=0$ yields 
$$
C_2/C_1=\mu_-/\mu_+.
$$
Hence,
\begin{equation}
B_n(\tau)=\mu_-\frac{1-e^{-\Delta_n\tau}}{1+(\mu_-/\mu_+)e^{-\Delta_n\tau}},
\label{a2}
\end{equation}
which is Eq. (\ref{B}). 

Plugging Eq. (\ref{a2}) into Eq. (\ref{A_def}) and setting $\xi=e^{-\Delta_ns}$ as a new integration variable we get
$$
A_n(\tau)=\frac{\theta\mu_-}{\Delta_n}\int_{e^{-\Delta_n\tau}}^1\frac{1-\xi}{\xi[1+(\mu_-/\mu_+)\xi]}d\xi,
$$
but
$$
\int\frac{1-\xi}{\xi[1+(\mu_-/\mu_+)\xi]}d\xi=\ln\xi-(1+\mu_+/\mu_-)\ln[1+(\mu_-/\mu_+)\xi].
$$
Hence (recall that $\mu_++\mu_-=\Delta_n$)
\begin{equation}
A_n(\tau)=\theta\left[\mu_-\tau+\ln\left(\frac{\mu_++\mu_-e^{-\Delta_n\tau}}{\Delta_n}\right)\right],
\label{a3}
\end{equation}
which is Eq. (\ref{A}).

\section{Behavior of the MET for small spans}
\label{appB}

Let us suppose $L\rightarrow 0$. From Eq. (\ref{delta}) we see that 
$$
\Delta_n=(\beta_n/L)\left[1+(L/\beta_n)^2/2+{\rm O}(L^4)\right]
$$
and
$$ 
\mu_{\pm}=(\beta_n/2L)\left[1\pm(L/\beta_n)+{\rm O}(L^2)\right].
$$
Hence
\begin{equation}
\frac{\mu_-^2}{\mu_+^2}=1-(4L/\beta_n)+{\rm O}(L^2),
\label{lambda_2}
\end{equation}
\begin{equation}
\frac{\mu_-}{\Delta_n}=[1-(L/\beta_n)+{\rm O}(L^2)]/2,
\label{lambda_delta}
\end{equation}
and 
\begin{equation}
\frac{1}{\mu_-}\left(\frac{\Delta_n}{\mu_+^2}\right)^\theta=
2^{2\theta+3}(L/\beta_n)^{\theta+1}\left[1+(1-2\theta)(L/\beta_n)+{\rm O}(L^2)\right].
\label{prefactor}
\end{equation}

(i) Suppose that $\theta<1$. Using Eq. (\ref{lambda_2}) we write
$$
F\left(\theta,\frac{\theta\mu_-}{\Delta_n};1+\frac{\theta\mu_-}{\Delta_n}; \frac{\mu_-^2}{\mu_+^2}\right)=F\left(\theta,\frac{\theta\mu_-}{\Delta_n};1+\frac{\theta\mu_-}{\Delta_n}; 1\right)+ {\rm O}(L),
$$
but \cite{mos}
\begin{equation}
F(a,b;c;1)=\frac{\Gamma(c)\Gamma(c-a-b)}{\gamma(c-a)\Gamma(c-b)}, \qquad(c-b-a>0).
\label{F1}
\end{equation}
Note that condition $c-b-a>0$ implies $\theta<1$. We thus find 
(cf. Eqs. (\ref{lambda_2})-(\ref{lambda_delta}))
\begin{equation}
F\left(\theta,\frac{\theta\mu_-}{\Delta_n};1+\frac{\theta\mu_-}{\Delta_n}; \frac{\mu_-^2}{\mu_+^2}\right)=\frac{\Gamma(1+\theta/2)\Gamma(1-\theta)}
{\Gamma(1-\theta/2)}[1+{\rm O}(L)], \qquad (\theta<1).
\label{approx_Fa}
\end{equation}
Plugging Eqs. (\ref{prefactor})-(\ref{approx_Fa}) into Eq. (\ref{exact_met}) and taking into account Eqs. (\ref{gamma}) and (\ref{theta}) we finally get
\begin{equation}
T(x)=N_1L^{\theta+1}\sum_{n=0}^\infty\frac{(-1)^n}{(2n+1)^{\theta+2}}
\cos\left[\left(2n+1\right)\pi x/L\right][1+{\rm O}(L)], \qquad(\theta<1),
\label{small_La}
\end{equation}
where
\begin{equation}
N_1=\left(\frac{2^{2\theta+5}}{\pi\alpha\theta}\right)\left(\frac{2\alpha}{\pi k}\right)^{\theta+1}
\frac{\Gamma(1+\theta/2)\Gamma(1-\theta)}{\Gamma(1-\theta/2)}.
\label{Na}
\end{equation}
For $x=0$ we have
\begin{equation}
T(0)= K_1 L^{\theta+1}[1+{\rm O}(L)], \qquad(\theta<1),
\label{T0a}
\end{equation}
where
$$
K_1=N_1\sum_{n=0}^\infty (-1)^n/(2n+1)^{\theta+2},
$$
and for small values of the span $L$ the MET grows as a power law with exponent related to the normal level of the volatility $\theta<1$. 

(ii) Suppose now that $\theta>1$. We employ the following property of the hypergeometric function \cite{mos}
$$
F(a,b;c;z)=(1-z)^{c-a-b}F(c-a,c-b;c;z),
$$
and write
$$
F\left(\theta,\frac{\theta\mu_-}{\Delta_n};1+\frac{\theta\mu_-}{\Delta_n}; \frac{\mu_-^2}{\mu_+^2}\right)=\left(1-\frac{\mu_-^2}{\mu_+^2}\right)^{1-\theta}
F\left(1-\theta+\frac{\theta\mu_-}{\Delta_n},1;\frac{\theta\mu_-}{\Delta_n}; \frac{\mu_-^2}{\mu_+^2}\right)
$$
which, after using Eq. (\ref{lambda_2}), reads
$$
F\left(\theta,\frac{\theta\mu_-}{\Delta_n};1+\frac{\theta\mu_-}{\Delta_n}; \frac{\mu_-^2}{\mu_+^2}\right)=\left(\frac{4L}{\beta_n}\right)^{1-\theta}
\left[F\left(1-\theta+\frac{\theta\mu_-}{\Delta_n},1;\frac{\theta\mu_-}{\Delta_n}; 1\right)
+{\rm O}(L)\right].
$$
Note that we can apply Eq. (\ref{F1}) since condition $c-a-b>0$ now implies $\theta>1$. Hence (cf. Eq. (\ref{lambda_delta}))
$$
F\left(\theta,\frac{\theta\mu_-}{\Delta_n};1+\frac{\theta\mu_-}{\Delta_n}; \frac{\mu_-^2}{\mu_+^2}\right)=\left(\frac{4L}{\beta_n}\right)^{1-\theta}
\left[\frac{1}{\theta-1}\Gamma(1+\theta/2)+{\rm O}(L)\right].
$$
Substituting this into Eq. (\ref{exact_met}) and taking into account Eq. (\ref{prefactor}) we obtain 
\begin{equation}
T(x)=N_2L^{2}\sum_{n=0}^\infty\frac{(-1)^n}{(2n+1)^{3}}
\cos\left[\left(2n+1\right)\pi x/L\right][1+{\rm O}(L)], \qquad(\theta>1),
\label{small_Lb}
\end{equation}
where
\begin{equation}
N_2=\left(\frac{2^{7}}{\pi\alpha\theta}\right)\left(\frac{2\alpha}{\pi k}\right)^{2}
\frac{\Gamma(1+\theta/2)}{1-\theta}.
\label{Nb}
\end{equation}
For $x=0$ we have
\begin{equation}
T(0)= K_2 L^{2}[1+{\rm O}(L)], \qquad(\theta>1),
\label{T0b}
\end{equation}
where
$$
K_2=N_2\sum_{n=0}^\infty (-1)^n/(2n+1)^{3}.
$$
Therefore, in this case the average escape time grows quadratically with the span --as if the zero-mean return would have followed the simple Brownian motion-- independently of the value of the normal level of volatility $\theta>1$. 

(iii) When $\theta=1$ we utilize the following series expansion of the hypergeometric function \cite{mos}
$$
F(a,b;a+b;z)=\frac{\Gamma(a+b)}{\Gamma(a)\Gamma(b)}\sum_{n=0}^\infty\frac{(a)_n(b)_n}{(n!)^2}]2\psi(n+1)-\psi(a+n)-\psi(b+n)-\ln|1-z|](1-z)^n,
$$
which, when $z\rightarrow 1$, yields the following approximation
$$
F(a,b;a+b;z)=-\frac{\Gamma(a+b)}{\Gamma(a)\Gamma(b)}\ln|1-z|+{\rm O}(1).
$$
Hence, as $L\rightarrow 0$,
$$
F\left(1,\frac{\mu_-}{\Delta_n};1+\frac{\mu_-}{\Delta_n}; \frac{\mu_-^2}{\mu_+^2}\right)=
-(1/2)\ln(4L/\beta_n)+{\rm O}(L\ln L),
$$
whence
\begin{equation}
T(x)=\frac{(4L)^2}{\alpha}
\sum_{n=0}^\infty\frac{\gamma_n}{\beta_n^2}\left[-\ln(4L/\beta_n)+
{\rm O}(L\ln L)\right]\cos\left[\left(2n+1\right)\pi x/L\right], \qquad(\theta=1),
\label{small_Lc}
\end{equation}
and there is a logarithmic growth with the span when $\theta=1$. 

The results above (cf. Eqs. (\ref{small_La}), (\ref{small_Lb}) and (\ref{small_Lc})) are summarized in 
Eq. (\ref{TL0}).

\section{Behavior of the MET for large spans}
\label{appC}

Let $T(x,y)$ be the MET of the joint process $(X(t),Y(t))$ out of the strip $-L/2\leq X(t)\leq L/2$, $0<Y(t)<\infty$. In terms of SP $S(x,y,t)$ the MET is given by
$$
T(x,y)=\int_0^\infty S(x,y,t)dt.
$$
From Eqs. (\ref{fpe})-(\ref{ibc}) we easily see that $T(x,y)$ is the solution to the boundary-value problem
\begin{equation}
\frac{1}{2}k^2y\frac{\partial^2 T}{\partial y^2}-\alpha(y-m^2)\frac{\partial T}{\partial y}+
\frac{1}{2}y\frac{\partial^2 T}{\partial x^2}=-1,
\label{met_eq}
\end{equation}
\begin{equation}
T(\pm L/2,y,t)=0.
\label{bc_met}
\end{equation}
Now our heuristic argument: large values of the span $L$ are equivalent to small values of $x$, but in this situation  (as long as $y$ is not too small) $\partial^2 T/\partial x^2$ is the dominant term in the left hand side of Eq. (\ref{met_eq}). This allows us to approximate the MET $T(x,y)\sim T_0(x,y)$, where the ``outer'' approximation \cite{bender} $T_0(x,y)$ is the solution to 
$$
\frac{1}{2}y\frac{\partial^2 T_0}{\partial x^2}=-1,\qquad T(\pm L/2,y,t)=0.
$$
That is, 
$$
T_0(x,y)=\frac{1}{y}\left[(L/2)^2-x^2\right],
$$
and the MET grows as
\begin{equation}
T(x,y)\sim L^2 \qquad (L\rightarrow\infty).
\label{parabolic}
\end{equation}


\begin{thebibliography}{99}
\bibitem{footnote1} We do not consider here the possible existence of cross correlation between the input noises $\xi_1(t)$ and $\xi_2(t)$. We will address in a future work the influence on the escape problem of a possible ``leverage'' correlation.
\bibitem{footnote2} The convergence of the numerical computation is rather fast and we get accurate enough results truncating the sum at $n=1000$ with an error tolerance below the $0.01\%$. The parameters of the model chosen are quite realistic in the sense that they are closer to those that capture the return probability density of Dow Jones daily data~\cite{yakov}. We have taken: $\theta=1.25$, $\alpha=0.045 \mbox{ day}^{-1}$, $m=0.093 \mbox{ day}^{-1/2}$ and $k=0.0014 \mbox{ day}^{-1}$ and we have thus considered the case with an interval $L=0.01$ and when $\tau=\alpha t=0.1$ ($t=2.22$ days).
\bibitem{footnote3} The numerical computation is in this case a bit more complicated than that of Fig. \ref{sp} since it requires the numerical evaluation of the integral~(\ref{t_n_v}) but in all situations we have found a rather fast convergence.
\bibitem{footnote4} In effect, the Fourier coefficient of $(2\alpha/k^2v)[(L/2)^2-x^2]+O(1/v^2)$ is precisely that of Eq. (\ref{t_n_infty_1}).
\bibitem{footnote_stat} This assumption is correct privided that the process started, for the first time, in the infinite past.
\end{thebibliography}
\end{document}